\documentclass[prd,nofootinbib,amsfonts]{revtex4}
\usepackage{graphicx,bm,amsmath,color}
\usepackage[latin1]{inputenc}

\newcommand{\be}{\begin{equation}}
\newcommand{\ee}{\end{equation}}

\begin{document}

\title{About Locality and the Relativity Principle Beyond Special Relativity}
\author{J.M. Carmona}
\email{jcarmona, cortes, dmazon, mercati@unizar.es}
\affiliation{Departamento de F\'{\i}sica Te\'orica,
Universidad de Zaragoza, Zaragoza 50009, Spain}
\author{J.L. Cort\'es}
\email{jcarmona, cortes, dmazon, mercati@unizar.es}
\affiliation{Departamento de F\'{\i}sica Te\'orica,
Universidad de Zaragoza, Zaragoza 50009, Spain}
\author{D. Maz\'on}
\email{jcarmona, cortes, dmazon, mercati@unizar.es}
\affiliation{Departamento de F\'{\i}sica Te\'orica,
Universidad de Zaragoza, Zaragoza 50009, Spain}
\author{F. Mercati}
\email{jcarmona, cortes, dmazon, mercati@unizar.es}
\affiliation{Departamento de F\'{\i}sica Te\'orica,
Universidad de Zaragoza, Zaragoza 50009, Spain}
\begin{abstract}
Locality of interactions is an essential ingredient of Special Relativity. Recently, a new framework
under the name of relative locality~\cite{AmelinoCamelia:2011bm} has been proposed as a way to
consider Planckian modifications of the relativistic dynamics of particles. We note in this
paper that the loss of absolute locality is a general feature of theories beyond Special
Relativity with an implementation of a relativity principle. We give an explicit construction
of such an implementation and compare it both with the previously mentioned framework of relative
locality and the so-called Doubly Special Relativity theories.
\end{abstract}

\maketitle

\section{Introduction}

Even if Special Relativity (SR) is at the basis of our present particle physics theories,
in the last years a strong motivation to explore departures of the Lorentz or
Poincaré invariance has emerged both from the theoretical and the phenomenological sides.
The idea that Lorentz invariance might be only
an approximate, low-energy, symmetry of Nature (as it happens with many other symmetries)
suggests the presence of corrections (suppressed by a certain high-energy scale) to the standard
relativistic dynamics (mainly through a modified dispersion relation and/or a modified
energy-momentum conservation law).

In fact, modified dispersion relations have arisen
in different approaches to quantum gravity such as: space-time foams~\cite{STF}, brane-world
scenarios~\cite{BW},  string theory~\cite{ST}, loop quantum
gravity~\cite{LQG}, renormalization group  of gravity~\cite{RG},
  analog models of gravity in condensed matter~\cite{AM}, canonical non-commutative~\cite{CNC} and $\kappa$-Minkowski~\cite{kM} spacetimes. Departures from Lorentz invariance have also been considered in several phenomenological contexts such as
proton decay in grand unification theories~\cite{Zee}, violations of the GZK cut-off~\cite{GZK}, anomalies in the cut-off of photons from Markarian 501~\cite{Mkn501}, neutrino oscillations~\cite{Osc} or tests of an energy-dependence of the speed of light with measurements of gamma ray bursts~\cite{STF}.

There are two main possibilities in a theory going beyond SR: either the relativity principle
is lost, which means there is a preferred reference frame (sometimes identified with that in which
the cosmic microwave background is isotropic), or it is preserved and the Poincaré invariance of SR is just \textit{deformed}. Concerning this last option it is important to stress that SR is not a consequence of only the relativity principle and the relativity of simultaneity, but also of the homogeneity of spacetime, the isotropy of space and some notion of causality, see \cite{Axiomatic}. We generally refer to the first case as an
scenario of Lorentz invariance violation (LIV). A particular example of it is
Kostelecky's Standard Model Extension~\cite{kostel}, in which Lorentz invariance is spontaneously
broken at low energies. The second case, however, implies the existence of a relativity principle
in the theory beyond SR. This was considered for the first time by G. Amelino-Camelia under the name
of Doubly Special Relativity (DSR)~\cite{DSRinic}.

Both scenarios have been mostly studied in momentum space, since a modified dispersion relation
(the relation between energy and momentum of a free particle) is usually the main ingredient
defining the extension or departure of SR. Interaction of particles may then be introduced through
an energy-momentum conservation law that may or may not show corrections with respect to the usual
additive law of SR (in the case of DSR, the conservation law must necessarily be modified in order
to be consistent with the relativity principle).

The space-time structure corresponding to these modifications of SR is however less known. Several
approaches to DSR suggest that it should contain noncommutative properties~\cite{DSRnoncomm}, although
a commutative spacetime has also been considered to be compatible with DSR models~\cite{DSRcomm}.
This lack of a well-defined space-time picture has caused some difficulties concerning, for example,
the correct expression for the velocity of a particle in terms of energy and
momentum~\cite{amelino_velocity,mignemi_velocity,aurelio_velocity,kowalski_velocity},
something which is crucial in order to study some experimental implications of the theory.

In fact, recently a debate about the phenomenological consistency of relativistic
theories beyond SR with an energy dependent velocity for photons
has emerged~\cite{debatesabine}. The key point in the discussion is the consequences of a loss of locality in the
theory. Actually, the issue of locality is also important in the case of a LIV scenario
(that is, without a relativity principle), in which the properties of translations in the underlying
spacetime and the Lorentz violating physics may
affect the energy-momentum conservation law as in the case of DSR~\cite{liv-conservationlaw}. As we
will see, there is an strict connection between the energy-conservation law and the space-time locality
of the interactions.

A byproduct of the above mentioned debate has been the introduction of a description of interactions in
spacetime with an associated new notion of `relative locality' in terms of a geometric interpretation
of the departures from SR kinematics~\cite{AmelinoCamelia:2011bm}. The proposal has been introduced in
a rather general way and, in its present state, a relativity principle has not been implemented yet.

One of the main ingredients of the relative locality paper Ref.~\cite{AmelinoCamelia:2011bm} is the emergence of spacetime from
momentum space through a variational principle. We think this idea represents a very important contribution
to the general problem of finding an appropriate space-time description in theories beyond SR. This
essential ingredient will be reviewed in Section II. Then, we will use this mathematical
construction to discuss the relationship between the notion of locality and the relativity principle
in a general framework with just two basic elements: the dispersion relation, describing the propagation
of free particles, and the energy-momentum conservation law, defining the interaction between these particles.
As we will see, an additive conservation law describes an interaction which is local in spacetime, and non-linear
corrections to this law causes the locality property to be lost for a general observer.
This means that the only way to preserve absolute locality
in a theory beyond SR is in a LIV scenario in which the dispersion relation is modified, but a linear
conservation law can be imposed. In particular, since a theory beyond SR with a relative principle (such as DSR)
requires a modification of both the dispersion relation and the conservation law, such a theory cannot be a local
theory.

In Section III we will consider the simplest implementation of a relativity principle, based on the use of the choice of appropriate phase space coordinates, in a theory in which locality is lost.
In Section IV we will discuss the physical meaning of the coordinates appearing in the model for the interaction of particles introduced in Section II and their relation to the notions of energy and momentum in a theory beyond SR.
The well-defined space-time description of interaction that emerges from the present work, generalizing
the SR image of interactions between particles as the crossing of space-time worldlines, will also allow us to give
answers to some paradoxes that have been pointed out in the context of DSR theories related with the velocity transformation law under boosts. A comparison of the present approach with that of DSR theories and with the geometric interpretation of Ref.~\cite{AmelinoCamelia:2011bm} will also be given in this Section.
We will end with a short summary and some concluding remarks in Section V, mentioning different alternatives to the simplest implementation of the relativity principle considered in this work.

\section{Interaction of particles in spacetime}

The description of a multiparticle process in spacetime is based on a
variational principle with an action
\begin{eqnarray}
S &=& \sum_{(-,a,i)} \int_{-\infty}^{s_a} ds
\left[x^{(i)\,\mu}_{-,a} \,
{\dot p}_{(i)\,\mu}^{\,-,a} + N_{-,a}^{(i)} \left[D(p^{\,-,a}_{(i)}) -
  m_{-,a}^{(i)2}\right]\right] \nonumber \\
&& + \sum_{(a,+,i)} \int_{s_a}^{\infty} ds \left[x^{(i)\,\mu}_{a,+}
{\dot p}_{(i)\,\mu}^{\,a,+} + N_{a,+}^{(i)} \left[D(p^{\,a,+}_{(i)}) -
  m_{a,+}^{(i)2}\right]\right] \nonumber \\
&& + \sum_{(a,b,i)} \int_{s_a}^{s_b} ds \left[x^{(i)\,\mu}_{a,b}
{\dot p}_{(i)\,\mu}^{\,a,b} + N_{a,b}^{(i)} \left[D(p^{\,a,b}_{(i)}) -
  m_{a,b}^{(i)2}\right]\right] \nonumber \\
&& + \sum_b z^{\mu}_b \left[K_{\mu}(p^{\,b,c}_{(i)},p^{\,b,+}_{(j)}) -
  K_{\mu}(p^{\,-,b}_{(i)},p^{\,a,b}_{(j)})\right](s_b)\,.
\label{action}
\end{eqnarray}
This action corresponds to a process with several ($V$) interactions,
each one characterized by the index $a=1, 2, ..., V$ and parameters
$s_a$ in increasing order $s_1 < s_2 < ... < s_V$. Phase space
coordinates for incoming particles are
($x^{(i)\,\mu}_{-,a}$, $p_{(i)\,\mu}^{\,-,a}$) and outgoing particle
coordinates ($x^{(i)\,\mu}_{a,+}$, $p_{(i)\,\mu}^{\,a,+}$). The index
$(-,a,i)$ ($(a,+,i)$) refers to the incoming (outgoing) particle $i$ in
(from) the $a$-interaction. There are also internal particles with phase
space coordinates ($x^{(i)\,\mu}_{a,b}$, $p_{(i)\,\mu}^{\,a,b}$) propagating
between interactions $a$ and $b$. There is a function $N$ for each particle
to implement the reparametrization invariance for the worldlines.
The function $D(p)$ determines the energy of a free particle of mass
$m$ and momentum ${\vec p}$ as the (positive) solution for $p_0$ of
the equation $D(p)=m^2$. There is also a constant vector $z^{\mu}_b$
to implement the energy-momentum conservation law in the
$b$-interaction
$K_{\mu}(p^{\,b,c}_{(i)},p^{\,b,+}_{(j)})=K_{\mu}(p^{\,-,b}_{(i)},p^{\,a,b}_{(j)})$.\footnote{We are considering the natural form for a conservation law with a separation of incoming and outgoing momentum variables. More general cases are discussed in the last section of the paper.}
The conservation law is defined by a set of functions, one for each
integer $n$,  $K: {\cal P}^n \to {\cal P}$ where ${\cal P}$ is the
four-momentum space of a particle and $K_{\mu}$ in the action
(\ref{action}) are the components of the image of these functions.

If one takes the addition for $K$ and $D(p)=p_0^2-\vec{p}^2$ then one
has the description of particle local interactions in special relativity
(SR). Other choices for $D$ and $K$ lead to generalizations of SR with
modified dispersion relations for the particles and non-linear
corrections in the energy-momentum conservation laws. Clearly,
these corrections require al least one dimensionful scale which would be
related to the Planck scale if these modifications come from quantum
gravitational effects.

The action~(\ref{action}), with one interaction, has been introduced
in Ref.~\cite{AmelinoCamelia:2011bm} in a discussion of the modification of
the notion of locality in a generalization of SR based on a new
understanding of the geometry of momentum space. The case of several
interactions has also been discussed in a particular case in Ref.~\cite{Freidel:2011mt}.
In these works the deviations from SR
kinematics are interpreted as a consequence of different
aspects of momentum space geometry with a metric geometry defined from
the function $D$ and non-linear corrections in conservation laws as a
consequence of a connection defined from the function $K$ for
$n=2$. This geometric interpretation of the generalization of SR
assumes that the functions $K$ with $n>2$ are obtained by iteration
from the case $n=2$ but we do not see any physical reason for this
additional assumption. From now on we will discuss the interaction of
particles and the issue of locality without any reference to a
(possible) geometric interpretation.

From the vanishing of the coefficients of the variations $\delta
x^{\mu}$, $\delta p_{\mu}$ for $s\neq s_a$, $\delta N$ and $\delta
z^{\mu}$ in the variation of the action we have the equations
\begin{equation}
{\dot p}_{(i)\,\mu}^{\,-,a} \,=\, {\dot p}_{(i)\,\mu}^{\,a,+} \,=\,
{\dot p}_{(i)\,\mu}^{\,a,b} \,=\, 0
\end{equation}
\begin{equation}
{\dot x}^{\,(i)\mu}_{-,a} \,=\, N_{-,a}^{(i)}
\frac{\partial D(p^{\,-,a}_{(i)})}{\partial p_{(i)\,\mu}^{\,-,a}}
{\hskip 1cm}
{\dot x}^{\,(i)\mu}_{a,+} \,=\, N_{a,+}^{(i)}
\frac{\partial D(p^{\,a,+}_{(i)})}{\partial p_{(i)\,\mu}^{\,a,+}}
{\hskip 1cm}
{\dot x}^{\,(i)\mu}_{a,b} \,=\, N_{a,b}^{(i)}
\frac{\partial D(p^{\,a,b}_{(i)})}{\partial p_{(i)\,\mu}^{\,a,b}}
\end{equation}
\begin{equation}
D(p^{\,-,a}_{(i)}) \,=\, m_{-,a}^{(i)2}
{\hskip 1cm}
D(p^{\,a,+}_{(i)}) \,=\, m_{a,+}^{(i)2}
{\hskip 1cm}
D(p^{\,a,b}_{(i)}) \,=\, m_{a,b}^{(i)2}
\label{dr}
\end{equation}
\begin{equation}
K_{\mu}(p^{\,b,c}_{(i)},p^{\,b,+}_{(j)}) \,=\,
  K_{\mu}(p^{\,-,b}_{(i)},p^{\,a,b}_{(j)}) {\hskip 1cm} b=1, 2, ..., V
\label{cl}
\end{equation}
i.e., the constancy of momentum, the relation between the four
velocity and momentum, the dispersion relation and the energy-momentum
conservation law at each interaction.

The vanishing of the coefficient of $\delta p_{\mu}$ for $s=s_a$ gives
the conditions
\begin{equation}
x^{(i)\,\mu}_{-,a} (s_a) \,=\, z_a^{\nu} \, \frac{\partial
  K_{\nu}(p^{\,-,a}_{(j)},p^{\,c,a}_{(k)})}{\partial p_{(i)\,\mu}^{\,-,a}}
{\hskip 1cm}
x^{(i)\,\mu}_{a,+} (s_a) \,=\,  z_a^{\nu} \, \frac{\partial
  K_{\nu}(p^{\,a,c}_{(j)},p^{\,a,+}_{(k)})}{\partial p_{(i)\,\mu}^{\,a,+}}
\end{equation}
which give the position of incoming and outgoing particles at the
interaction. When $K$ deviates from the addition one has different
particles at different points (non-local interactions) as a
consequence of the non-linear corrections to the energy-momentum
conservation laws. These conditions, together with the equation which
gives the four-velocity in terms of the momentum, fix the
worldlines that describe the propagation of the incoming
particles before the interaction and the outgoing particles after the
interaction:\footnote{As usual, we make use of the reparametrization invariance of worldlines to choose $s$ such that $N$ is constant.}
\begin{equation}
x^{(i)\,\mu}_{-,a} (s) \,=\, z_a^{\nu} \, \frac{\partial
  K_{\nu}(p^{\,-,a}_{(j)},p^{\,c,a}_{(k)})}{\partial p_{(i)\,\mu}^{\,-,a}}  \,+\,
(s - s_a) \, N_{-,a}^{(i)} \,
\frac{\partial D(p^{\,-,a}_{(i)})}{\partial p_{(i)\,\mu}^{\,-,a}}
\label{wl-a}
\end{equation}
\begin{equation}
x^{(i)\,\mu}_{a,+} (s) \,=\, z_a^{\nu} \, \frac{\partial
  K_{\nu}(p^{\,a,c}_{(j)},p^{\,a,+}_{(k)})}{\partial p_{(i)\,\mu}^{\,a,+}} \,+\,
(s - s_a) \, N_{a,+}^{(i)} \,
\frac{\partial D(p^{\,a,+}_{(i)})}{\partial p_{(i)\,\mu}^{\,a,+}}\,.
\label{wla+}
\end{equation}
For the coordinates of the particles propagating between interactions
$a$ and $b$ one has two conditions from the vanishing of the coefficients
of $\delta p_{(i)\,\mu}^{\,a,b}$  at $s=s_a$ and $s=s_b$
\begin{equation}
x^{\,(i)\mu}_{a,b} (s_a) \,=\, z_a^{\nu} \, \frac{\partial
  K_{\nu}(p^{\,a,c}_{(j)},p^{\,a,+}_{(k)})}{\partial
  p_{(i)\,\mu}^{\,a,b}}
{\hskip 1cm}
x^{\,(i)\mu}_{a,b} (s_b) \,=\,  z_b^{\nu} \, \frac{\partial
  K_{\nu}(p^{\,-,b}_{(j)},p^{\,c,b}_{(k)})}{\partial p_{(i)\,\mu}^{\,a,b}}\,.
\end{equation}
One of these conditions can be used to determine the worldline of the
particle propagating between the two interactions
\begin{equation}
x^{\,(i)\mu}_{a,b} (s) \,=\, z_a^{\nu} \, \frac{\partial
  K_{\nu}(p^{\,a,c}_{(j)},p^{\,a,+}_{(k)})}{\partial
  p_{(i)\,\mu}^{\,a,b}} \,+\, (s - s_a) \, N_{a,b}^{(i)} \,
\frac{\partial D(p^{\,a,b}_{(i)})}{\partial p_{(i)\,\mu}^{\,a,b}}
\label{wlab}
\end{equation}
but there is one extra condition
\begin{equation}
z_b^{\nu} \, \frac{\partial
  K_{\nu}(p^{\,-,b}_{(j)},p^{\,c,b}_{(k)})}{\partial
  p_{(i)\,\mu}^{\,a,b}} -
z_a^{\nu} \, \frac{\partial
  K_{\nu}(p^{\,a,c}_{(j)},p^{\,a,+}_{(k)})}{\partial
  p_{(i)\,\mu}^{\,a,b}} \,=\, (s_b - s_a) N_{a,b}^{(i)}
\frac{\partial D(p^{\,a,b}_{(i)})}{\partial p_{(i)\,\mu}^{\,a,b}}
\label{zz}
\end{equation}
to be taken into account when one identifies the general solution of
the variational principle based on the action (\ref{action}).

Any choice of momenta compatible with the dispersion relations
(\ref{dr}) and the conservation laws (\ref{cl}) and any choice of
coordinates $z_a$ for the interactions compatible with the conditions
(\ref{zz}), which can also restrict the possible choice of momenta,
define a solution for the space-time description of the multiparticle
process. The general solution can be obtained by considering a set of $V-1$ internal lines such that they connect all of the $V$ vertices (this is known as a ``minimal tree''). One can first use $V-1$ of the constraints (\ref{cl}) to fix the internal momenta of the minimal tree in terms of the external momenta and the $I-V+1$ remaining internal momenta outside the minimal tree. Next one can take the $V-1$ constraints in Eq.~(\ref{zz}) corresponding to the momenta in the minimal tree as a set of linear equations for $V-1$ of the coordinates $z_a$ whose solution gives these coordinates in terms of one remaining coordinate, for example $z_1$, and the momenta. The remaining $I-V+1$ constraints in Eq.~(\ref{zz}) can then be used to determine the internal momenta outside the minimal tree as a function of $z_1$ and the external momenta. Finally there is one constraint in Eq.~(\ref{cl}) which gives a condition for the external momenta.

The fact that there is a solution for each choice of $z_1^\mu$ implies the translational invariance of the model: in fact, given a solution,
one can obtain a 4-parameter family of equivalent solutions, given by
\begin{equation}
z'_a = z_a + \delta z_a ~, \label{translations}
\end{equation}
where the $V$ parameters $ \delta z_a$ satisfy $(V-1)$ independent equations,
\begin{equation}
\delta z_b^{\nu} \, \frac{\partial
  K_{\nu}(p^{\,-,b}_{(j)},p^{\,c,b}_{(k)})}{\partial
  p_{(i)\,\mu}^{\,a,b}} -
\delta z_a^{\nu} \, \frac{\partial
  K_{\nu}(p^{\,a,c}_{(j)},p^{\,a,+}_{(k)})}{\partial
  p_{(i)\,\mu}^{\,a,b}} \,=0 ~.
\end{equation}
The above equations determine all but one of the $\delta z_a$, for example $\delta z_1$: there are
4 independent translations.

The transformations (\ref{translations}) connecting solutions with different choices for $\delta z_1$ reflect the absence of any reference to a point in the space-time description of the multiparticle process. These solutions correspond to the description of the process by different observers related by translations. Translations are defined then as transformations on the set of worldlines of the particles participating in the process. The transformation of each worldline depends on the momenta of different particles and on the proper time intervals $s_b-s_a$. This nontrivial realization of translations, in contrast with a space-time translation of each of the points of the different worldlines, is a direct consequence of the nonlinear corrections in the energy-momentum conservation laws. There is an observer for which $z_1=0$ and then all the incoming and outgoing  particles in one of the interactions are at the origin of the coordinates of spacetime. This is the reason why one refers to this framework as a replacement of absolute locality by relative locality~\cite{AmelinoCamelia:2011bm}. Any physical process always involves more than one interaction (in the simplest case one has an interaction in the production of a particle and an interaction in its detection) and then one can say that in the presence of non-linear corrections in the energy-momentum conservation laws one has real, physical consequences of the loss of absolute locality. However in order to have observable consequences of the loss of absolute locality one requires a process with interactions separated by very long distances~\cite{Freidel:2011mt}.

\section{Relativity principle}

In the case of SR ($D(p)=p_0^2-\vec{p}^2$ and $K(\{p\})=\sum p$) there are, together with translations, also transformations among solutions with different incoming and outgoing momenta. Given a solution ($\{p\}$, $z$) of the variational principle for the action (\ref{action}) in SR, a combination ($\{p'\}$, $z'$) with
\begin{equation}
p'_{\mu} = {\Lambda_{\mu}}^{\nu} \,\, p_{\nu} {\hskip 2cm}
z'^\mu = \,\eta^{\mu\rho} {\Lambda_{\rho}}^{\sigma} \eta_{\sigma\nu} \, z^{\nu}
= {\Lambda ^{\mu}}_{\rho} \, \, z^{\rho}
\end{equation}
will also be a solution if ${\Lambda_{\mu}}^{\nu}$ is a Lorentz transformation (where we used that $\eta^{\rho\sigma}=\eta^{\mu\nu}{\Lambda_\nu}^\rho {\Lambda_\mu}^\sigma=(\Lambda^T\eta\Lambda)^{\rho\sigma}$).

In the general case one can always make use of canonical transformations in order to study the properties of the solutions of the variational principle. We will consider a function $D(p)$ in the modified dispersion relation such that it is possible to find new momentum variables $\pi_{\mu}$, defined by a set of non-linear functions $F_{\mu}$ through $\pi_{\mu}\equiv F_{\mu}(p)$, so that
\begin{equation}
\pi_0^2-\vec{\pi}^2 \,=\, \left[F_0(p)\right]^2 - \sum_{i=1}^3 \left[F_i(p)\right]^2 \,=\, D(p).
\label{auxvars}
\end{equation}
The energy-momentum conservation laws in terms of the new momentum variables require to consider
\begin{equation}
K_\mu(\{F^{-1 }(\pi)\})\equiv {\cal K}_\mu(\{\pi\}).
\label{calK}
\end{equation}
In order to identify transformations among solutions of the variational principle it is necessary to identify transformations of the momentum variables compatible with the restrictions on those variables from the dispersion relation and the conservation laws. The dispersion relation is invariant under a Lorentz transformation
\begin{equation}
\pi_{\mu}'^{\, (i)} = {\Lambda_{\mu}}^{\nu} \,\, \pi_{\nu}^{(i)}
\label{piLorentz}
\end{equation}
of the new momentum variables. If the conservation laws are such that
\begin{equation}
{\cal K}_{\mu} (\{\pi\}) \,=\, \sum_{i=1}^{N} f_i(\{\pi\}) \pi_{\mu}^{(i)}
\label{RP}
\end{equation}
with $f_i(\{\pi'\})=f_i(\{\pi\})$, i.e. a function of the $N(N+1)/2$ Lorentz invariant combinations of the $N$ momenta $\{\pi\}$, then the conservation laws will also be invariant under the transformation (\ref{piLorentz}) and then ($\{p'\}$, $z'$) with
\begin{equation}
p'_{\mu} \,=\, F_{\mu}^{-1}(\pi')  {\hskip 2cm}
z'^\mu = \,\eta^{\mu\rho} {\Lambda_{\rho}}^{\sigma} \eta_{\sigma\nu} \, z^{\nu}
= {\Lambda ^{\mu}}_{\rho} \, \, z^{\rho}
\label{pzLorentz}
\end{equation}
will be a solution of the variational problem if ${\Lambda_{\mu}}^{\nu}$ is a Lorentz transformation and if ($\{p\}$, $z$) is a solution. In this case we have a space-time description of a multiparticle process going beyond SR and compatible with a relativity principle.

Together with the new momentum variables one can introduce new space-time coordinates $\xi$ through
\begin{equation}
x^{\mu} \,=\, \xi^{\nu} \, \frac{\partial F_{\nu}}{\partial p_{\mu}}
\label{xi}
\end{equation}
such that $x^{\mu} {\dot p}_{\mu} = \xi^{\nu} {\dot \pi}_{\nu}$ (i.e., to complete a canonical transformation). The variational problem can be solved in terms of the variables $\xi$, $\pi$ and the solution for the worldlines of the different particles will be
\begin{equation}
\xi^{(i)\,\mu}_{-,a} (s) \,=\, z_a^{\nu} \, \frac{\partial
  {\cal K}_{\nu}(\pi^{\,-,a}_{(j)},\pi^{\,c,a}_{(k)})}{\partial \pi_{(i)\,\mu}^{\,-,a}}  \,+\,
(s - s_a) \, N_{-,a}^{(i)} \, 2 \eta^{\mu\nu} \pi_{(i)\,\nu}^{\,-,a}
\label{xiwl-a}
\end{equation}
\begin{equation}
\xi^{(i)\,\mu}_{a,+} (s) \,=\, z_a^{\nu} \, \frac{\partial
  {\cal K}_{\nu}(\pi^{\,a,c}_{(j)},\pi^{\,a,+}_{(k)})}{\partial \pi_{(i)\,\mu}^{\,a,+}} \,+\,
(s - s_a) \, N_{a,+}^{(i)} \,  2 \eta^{\mu\nu} \pi_{(i)\,\nu}^{\,a,+}
\label{xiwla+}
\end{equation}
\begin{equation}
\xi^{\,(i)\mu}_{a,b} (s) \,=\, z_a^{\nu} \, \frac{\partial
  {\cal K}_{\nu}(\pi^{\,a,c}_{(j)},\pi^{\,a,+}_{(k)})}{\partial
  \pi_{(i)\,\mu}^{\,a,b}} \,+\, (s - s_a) \, N_{a,b}^{(i)} \,
2 \eta^{\mu\nu} \pi_{(i)\,\nu}^{\,a,b}\,.
\label{xiwlab}
\end{equation}
When the conservation laws (\ref{RP}) are chosen in a way compatible with the relativity principle then one has that the Lorentz transformations connecting different solutions of the variational principle act on the worldlines as
\begin{equation}
\xi'^\mu (s) \,=\, \eta^{\mu\nu} {\Lambda_{\nu}}^{\sigma} \eta_{\sigma\rho} \,\xi^{\rho} (s)
= {\Lambda ^{\mu}}_{\rho} \, \, \xi^{\rho} (s)
\label{xitransform}
\end{equation}
which is the Lorentz transformation law of worldlines in special relativity.
This transformation of worldlines can be reexpressed in terms of the original space-time coordinates by using Eq.~(\ref{xi}).

The simplest example of a conservation law compatible with a relativity principle corresponds to the case of two particles ($N=2$) and one has
\begin{equation}
{\cal K}_{\mu} (\pi^{(1)}, \pi^{(2)}) \,=\, f(\pi^{(11)}, \pi^{(12)}, \pi^{(22)}) \, \pi_{\mu}^{(1)} \,+\, f(\pi^{(22)}, \pi^{(12)}, \pi^{(11)})  \, \pi_{\mu}^{(2)}
\label{twopartconslaw}
\end{equation}
with $\pi^{(ij)} \,=\, \eta^{\mu\nu} \pi^{(i)}_{\mu} \pi^{(j)}_{\nu}$.
The general form of the conservation law is determined by a function $f$ of three variables. If one assumes that the conservation laws with more than two particles are obtained from the case $N=2$ by iteration then the function $f$ determines all conservation laws. It has been shown~\cite{AmelinoCamelia:2011bm} that in this case it is possible to define a connection on momentum space related to $f$; the implementation of the relativity principle as discussed in this Section leads to a geometry of momentum space which is fixed by the functions $D$ and $f$.
In the general case the conservation law with more than two particles involves new independent functions and there is no geometrical interpretation of the generalization of SR.

\section{Physical interpretation and relation with other proposals}

What is the physical meaning of the momentum coordinates $p_\mu$ and the space-time coordinates $x^\mu$ used in the description of the multiple-particle process? The energy-momentum of a particle has to be determined through a measurement process, which gives certain values $p_\mu$. This is done using apparatuses that are calibrated in such a way that they measure the special relativistic energy and momentum in the limit where corrections to SR can be neglected. However, when these corrections are taken into account, we do not know exactly what our apparatus is measuring and in fact different apparatuses may give different results.

In the simplest kinematic generalization of SR one considers a modification of the dispersion relation at very high energies but the usual additive energy-momentum conservation laws. This is included in the general framework based on the action (\ref{action}) with $K$ additive but a nontrivial function $D$. In this case one has a violation of Lorentz invariance but locality is preserved.

When one goes beyond SR in a way compatible with the relativity principle then one has the action (\ref{action}) with
\be
D(p) \,=\, \left[ F_{0}(p)\right] ^2 - \sum_{i=1}^3 \left[F_i(p)\right]^2
\ee
and a nonlinear energy-momentum conservation law fixed by
\be
K_{\mu}(\{p\}) \,=\, {\cal K}_{\mu}(\{F(p)\})
\label{K(p)}
\ee
defined in terms of the ${\cal K}_{\mu}$ introduced in Eq.~(\ref{RP}). Absolute locality is lost in this case owing to the nonlinearity of $K_{\mu}(\{p\})$, which arises in principle from two sources: the functions $F$ and ${\cal K}$. The implementation of the relativity principle and in particular the generalized boost transformation of energy-momentum is fixed by the nonlinear mapping in momentum space $F$. Different results for different measurement apparatuses would correspond to a dependence of the mapping $F$ on the choice of the energy-momentum measurement apparatus. The coordinates $\pi_{\mu} = F_\mu (p)$ are familiar in the context of generalizations of the relativity principle (DSR) compatible with an observer-independent high-energy scale~\cite{class,judes}. They are considered as a useful tool in DSR under the name of \emph{auxiliary} (or classical) energy-momentum variables. However, in canonical implementations of DSR the auxiliary variables are normally considered to compose additively~\cite{judes},\footnote{The fact that this does not need to be necessarily the case was indicated for the first time in Ref.~\cite{asr}.} and in this case all the departures from SR including the loss of absolute locality would be due to the difference between the results of energy-momentum measurements (the $p_\mu$ variables) and the auxiliary variables $\pi_\mu$. Different choices for the mapping $F$ depending on a new energy scale (to be identified with the Planck mass if one assumes a gravitational origin for the generalization of SR) can reproduce the different versions of DSR. Nevertheless, in the general case one has nonlinear corrections in the energy-momentum conservation laws both from the mapping $F$ and from ${\cal K}_{\mu}$ in Eq.~(\ref{RP}). This gives a new perspective on DSR with a different realization of spacetime and invariance under translations. In fact the auxiliary space-time coordinates $\xi^\mu$ corresponding to the auxiliary energy-momentum variables $\pi_\mu$ share an
important property with the spacetime of SR: it is possible to speak of their transformed space-time coordinates under a Lorentz boost (see Eq.~(\ref{xitransform})). This is not so, however, with respect to translations, as it can be seen from Eqs.~(\ref{xiwl-a})-(\ref{xiwlab}) (recall that a translation has been defined as $z_1^\nu\to z_1^\nu +\epsilon^\nu$, where $z_1$ are the coordinates of one of the interactions).
If one takes the measured energy-momentum $p_\mu$ and the corresponding spacetime $x^\mu$ then one has to consider also boost transformations on phase space.

Some paradoxes that have been pointed out in the context of DSR can be solved in the present perspective of generalizations of special relativity. Specifically, it has been indicated that the `natural' definition of velocity in DSR, as in Hamiltonian mechanics with a canonical phase-space, $v_g\equiv \partial E/\partial p$~\cite{mignemi_velocity} (which also allows one
to interpret it as the group velocity of wave packets representing point particles in specific
DSR frameworks~\cite{amelino_velocity}), poses some interpretational problems of the velocity as the parameter of Lorentz transformations
(because the relation between relative velocity and rapidity depends on the mass of the particle), and with
the principle of relativity (two particles which interact in a reference frame (RF) may never meet, as
seen from another RF)~\cite{mignemi_velocity,aurelio_velocity}.
A different relation between the velocity and the energy-momentum variables arises in the context of free point particles propagating in $\kappa$-Minkowski spacetime, in such a way that $v\neq v_g$ satisfies the usual relativistic composition law.
This avoids the previous paradoxes at the cost of giving up the relation between
wave packets and point particles in these theories~\cite{kowalski_velocity}.

The main difficulty to define velocity in DSR theories is that, being formulated in momentum space, they miss a clear space-time picture that should
give an unequivocal answer. In the present work, however, one can explicitly calculate the velocity as
\begin{equation}
v^j\equiv\frac{\dot x^j}{\dot x^0}=\left. \frac{\partial D/\partial p_j}{\partial D/\partial p_0}\right|_{D(p)=m^2}=\left.\frac{d E}{d p_j}\right|_{E=p_0(\vec{p},m^2)},
\label{veqdEdp}
\end{equation}
so that indeed $v=v_g$. If one forgets the difference between the auxiliary energy-momentum variables ($\pi_\mu$) and the results of the energy-momentum measurement ($p_\mu$) one would have a velocity
\begin{equation}
\upsilon^j=\frac{\dot\xi^j}{\dot\xi^0}=\left.\frac{d \epsilon}{d \pi_j}\right|_{\epsilon=\sqrt{\vec\pi^2+m^2}}=\frac{\pi _j}{\epsilon},
\label{veqpieps}
\end{equation}
and then a composition law for velocities which is exactly that of SR, so that no paradoxes arise. If however one uses the physical phase-space variables ($x^{\mu}$, $p_{\mu}$), then the transformation of the velocity of a particle between two inertial observes will indeed depend on the mass of the particle~\cite{mignemi_velocity}. Nevertheless, there is no contradiction with the relativity principle. The argument saying that two particles of different masses which are at relative rest for one system of reference, may interact (will necessarily do so in 1$+$1 dimension) for another observer is based on the naive (special-relativistic) picture that the interactions are given by the crossing of worldlines. When absolute locality is broken, as it happens in a theory beyond SR with a relativity principle, this picture is not right any more, and two worldlines that are parallel for one observer in 1$+$1 dimension will in general interact. The fact that two worldlines which interact for one observer will necessarily interact for any other inertial observer is automatically incorporated into the formalism presented in the previous Section, since the transformations between observers have been defined as transformations between solutions of the variational problem defined in Section II. Even if the new transformation laws of the velocities in Eq.~(\ref{veqdEdp}) do not lead to any contradiction, it is possible to find particular cases where in fact velocities transform as in SR. An example is given by a model with
\be
F_0 (p) \,=\, \dfrac{p_0 - \lambda \, \vec{p}^2}{\left( 1- \lambda \, p_0 \right) \sqrt{1- \lambda ^2 \, \vec{p}^2}}\,, {\hskip 2cm}
F_i (p) \,=\, \dfrac{p_i}{\sqrt{1- \lambda ^2 \, \vec{p}^2}}\,,
\ee
which gives the dispersion relation of the Magueijo-Smolin DSR proposal~\cite{MAG02} $D(p)=\left({p_0}^2-\vec{p}^2\right)/\left( 1-\lambda \, p_0\right) ^2$, with $\lambda$ the typical  length scale of DSR. It is easy to see that one has in this case
\be
v^i \, = \, \frac{\partial D/\partial p_i}{\partial D/\partial p_0} \, = \,  \, \dfrac{ p_i \left( 1-\lambda \, p_0 \right)}{p_0 - \lambda \, \vec{p}^2} \, = \, \frac{\pi_i}{\pi_0}\,,
\ee
and then velocities transform as in SR. This is the case independently of the choice of the energy-momentum conservation law (choice of ${\cal K}_{\mu}$). Examples with the dispersion relation of other DSR proposals and with special-relativistic velocity transformation laws can also be easily formulated.

The proposal $v=v_g$ also produces an energy dependent velocity of photons in certain DSR models, whose consistency
with (macroscopic) experimental observations has been recently questioned and debated~\cite{debatesabine}. Neglecting the difference between auxiliary energy-momentum variables and the measured energy-momentum, the velocity of photons is independent of the energy. This is also the case if $F$ is such that the model has a special relativistic velocity transformation law. In any case the real issue is not the energy dependence or independence of the velocity of photons but the consequences of the absolute locality breaking, and there is a consistency with experimental observations because all of them correspond to an observer whose origin is not far away from the observations corresponding to interactions, which are well approximated by an absolutely local interaction model.

The possibility that the measurement process could define different `physical' momenta depending on the apparatus is connected with the freedom in the choice of `calorimeters' that has been introduced in Refs.~\cite{AmelinoCamelia:2011bm,Freidel:2011mt,AmelinoCamelia:2011uk}, in which the proposed generalization of SR has a reading in terms of a geometry of the momentum space.

A geometry in momentum space can always be introduced if one considers a generalized energy-momentum conservation law which can be derived from the conservation law for a two particle system. Taking auxiliary variables $\pi_\mu$ as coordinates in momentum space, a connection can be obtained from the algebra associated to the nonlinear conservation law ${\cal K}_\mu(\{\pi\})$ in Eq.~(\ref{twopartconslaw}) following the general procedure introduced in Ref.~\cite{AmelinoCamelia:2011bm}. An internal law $\oplus$ is defined in momentum space by
\begin{equation}
\left[\pi^{(1)} \oplus \pi^{(2)}\right]_\mu \,=\, {\cal K}_{\mu} (\pi^{(1)}, \pi^{(2)})
\label{pluspi}
\end{equation}
and the inverse $\ominus\pi$ by
\begin{equation}
{\cal K}_{\mu} (\pi, \ominus\pi) \,=\, 0\,
\end{equation}
where it is necessary to assume that $f(x, 0, 0)=f(0, 0, x)=1$ so that the origin in momentum space is the identity of the composition law.
In the case of a function $f$ symmetric under exchange of its first and third arguments, the inverse takes the simple form $[\ominus\pi]_\mu = - \pi_\mu$ and the result for the connection can be written in a compact form,
\be
\Gamma^{\beta\gamma}_\alpha (\pi) \,=\, - \left[f(\pi^2, -\pi^2, \pi^2) \, f(0, 0, 0)\right]^2 \, \left[\left(\delta_\alpha^\beta \pi^\gamma + \delta_\alpha^\gamma \pi^\beta\right) \partial_2 f(\pi^2, 0, 0) + 2 \eta^{\beta\gamma} \pi_\alpha \partial_3 f(\pi^2, 0, 0) + \pi_\alpha \pi^\beta \pi^\gamma \partial_2\partial_2 f(\pi^2, 0, 0)\right],
\ee
where $\partial_2$ ($\partial_3$) denotes the partial derivative of the function $f$ of three variables in Eq.~(\ref{twopartconslaw}) with respect to its second (third) argument.
It is clear that one does not have a torsion as a consequence of the commutativity of the algebra associated to the conservation law defined by ${\cal K}_\mu$ in Eq.~(\ref{twopartconslaw}). 

If one uses the measured energy-momentum $p_\mu$\footnote{Or, in geometric language, if one changes coordinates in momentum space.} then the internal law in Eq. (\ref{pluspi}) takes the form
\begin{equation}
\left[p^{(1)} \oplus p^{(2)}\right]_\mu \,=\, {\tilde {\cal K}}_{\mu} (p^{(1)}, p^{(2)})
\end{equation}
with 
\begin{equation}
{\tilde {\cal K}}_\mu (\{p\}) \,=\, F^{-1}_\mu \left({\cal K}(\{F(p)\})\right) \,=\, F^{-1}_\mu \left(K(\{p\})\right)\,,
\end{equation}
where in the second equality we have used Eq.~(\ref{K(p)}).
The connection in the coordinates $p_\mu$ is 
\be
\tilde{\Gamma}^{\nu\rho}_\mu (p) \,=\, \left[\frac{\partial F_\alpha}{\partial p_\mu} \frac{\partial F_\beta}{\partial p_\nu} \Gamma^{\beta\gamma}_\alpha \left(F(p)\right) + \frac{\partial^2 F_\gamma}{\partial p_\mu \partial p_\nu}\right] \,\frac{\partial F_\rho^{-1}}{\partial \pi_\gamma}\left(F(p)\right)\,.
\label{Gammatilde}
\ee

We have included the expression for the connection in Eq.~(\ref{Gammatilde}) to make explicit the geometric reformulation
of the implementation of the relativity principle proposed in Section III and its relation to the proposal of relative locality~\cite{AmelinoCamelia:2011bm} but we do not think that this is relevant for an analysis of the phenomenological consequences of the generalization of SR proposed in this work.

In the geometric context one considers observables which can be expressed in terms of intrinsic geometric properties~\cite{Freidel:2011mt}. The interpretation we are proposing is however less restrictive. It is possible to find special cases of observables whose expression in terms of the coordinates do not depend on the choice of coordinates, and in the particular case when there is a geometrical interpretation for the generalization of SR these observables will have a geometrical meaning. Their values will be independent of the choice of the energy-momentum measurement apparatus. But surely this is not a necessary condition that an observable should fulfil: the energy-momentum in SR (and its generalization) is a counterexample. In this sense the difference between our interpretation of variables and observables and that of Refs.~\cite{AmelinoCamelia:2011bm,Freidel:2011mt,AmelinoCamelia:2011uk} is similar to the one existing between Special Relativity (which is specifically contained in our proposal) and General Relativity.

In the discussion of the phenomenological implications of a modification of SR one requires to establish a correspondence among observables. In general one finds different possible candidates (different measurement apparatuses) for a given observable in SR. This ambiguity puts limitations on the predictability of the general framework used to discuss generalizations of SR. The possibility to find in some cases a candidate with a geometrical meaning does not eliminate the limitations on the predictability. The selection of this candidate is a choice among different observables. An application of these arguments in the example of the energy-dependence of the time delay of different signals from sources at very long distances will be presented elsewhere~\cite{loc-RP-ph}.

\section{Concluding remarks}

We have presented a framework describing the interactions of particles in spacetime in a theory beyond SR. We have seen that a relativity principle (with a generalization of both translations and Lorentz transformations) may be implemented in the theory, which inevitably causes a loss of absolute locality. The central idea of describing interactions in spacetime as a generalization of the crossing of worldlines in SR through a variational principle was introduced in Ref.~\cite{AmelinoCamelia:2011bm}. We describe however a more general framework which does not necessarily include the geometric interpretation of the formalism defined in Ref.~\cite{AmelinoCamelia:2011bm} (for which a detailed discussion of the implementation of a relativity principle is still missing). We have also considered the physical interpretation of the variables appearing in the variational problem and the similarities and differences with DSR. In particular, some paradoxes concerning the definition of velocity in DSR find their natural solution within the present work. Phenomenological implications of the present way to introduce interactions beyond SR, such as time-delay calculations, will be presented elsewhere~\cite{loc-RP-ph}.

At this point we would like to make some comments on the generality of the implementation of the relativity principle based on the functions ${\cal K}_{\mu}$ in Eq.~(\ref{RP}). When imposing ${\cal K}_{\mu} (\{\pi\}^{\Lambda})={\Lambda_{\mu}}^{\nu} {\cal K}_{\nu} (\{\pi\})$, we have implicitly assumed that under a Lorentz transformation
\be
\{\pi\}^{\Lambda} = \{\pi^{\Lambda}\},
\ee
i.e., we have assumed that under a Lorentz transformation the set of momenta  of different particles transform independently.
But we could consider more general cases where one has directly a realization of Lorentz transformations on the set of momenta of the particles in each interaction. In fact we have seen that a translation requires to consider all the worldlines of the particles in each interaction and one does not have a realization of translations acting independently on each worldline. It could be interesting to explore these alternatives to the implementation of the relativity principle and their possible relation with a noncommutativity in spacetime and nontrivial implementations of boost transformations for multiparticle states.

There is still another possibility to try to implement a relativity principle in a generalization of SR. As discussed in Ref.~\cite{AmelinoCamelia:2011bm}, the idea of relative locality has a natural realization within a perspective based on a geometry of the momentum space. From this point of view it is natural to try to implement the relativity principle through the isometries of this geometry. In this way one does not need to make a choice of energy-momentum measurement apparatus as a starting point for the implementation of the relativity principle. It could be interesting to see the relation between this geometric implementation of the relativity principle and the simplest implementation based on the functions ${\cal K}_{\mu}$ in Eq.~(\ref{RP}) or the alternatives with a Lorentz transformation on the whole set of momenta of the particles participating in each interaction.

Another remark on the generality of the proposal for a generalization of SR based on the action (\ref{action}) is that we could have considered more general forms of the energy-momentum conservation laws. From the algebra associated to the composition of two momenta it is possible to define the inverse of a given momentum. One possibility to implement the conservation laws is to consider the vanishing of the momentum obtained by successive composition of the momenta of the incoming particles and the inverse of the momenta of the outgoing particles. Another possibility, the one we have considered in our proposal, is based on the identification of an initial momentum as a result of successive compositions of the momenta of the incoming particles and a final momentum resulting from the composition of the outgoing particles. The conservation laws amount to the equality of the initial and final momentum or, equivalently, to the vanishing of the composition of the initial momentum and the inverse of the final momentum.

As a final remark, all the discussion in this work is based on a classical model for multiparticle interactions based on the action (\ref{action}). It seems also interesting to explore the possibility to include quantum effects by going beyond the variational principle and considering a sum over paths weighted with the exponential of the action (\ref{action}) with just one interaction.\footnote{It is not necessary to include several interactions in the action; the contribution of processes with several interactions is included in the expansion of the exponential.} This could correspond to a first-quantized approach instead of a field theoretical approach to the generalized quantum theory.

\section*{Acknowledgments}
This work is supported by CICYT (grant FPA2009-09638) and DGIID-DGA (grant
2010-E24/2). D.M. acknowledges a FPI grant from MICINN.



\end{document}